\documentclass[letterpaper, 10 pt, conference]{ieeeconf}  

\IEEEoverridecommandlockouts                              

\usepackage{amsmath}
\usepackage{cite} 
\usepackage{hyperref}
\usepackage{graphicx}
\usepackage{subcaption}
\graphicspath{{./pdf/}{./jpeg/}{./img/}}
\title{\LARGE \bf
Comparative Analysis of CNN and Transformer Architectures with Heart Cycle Normalization for Automated Phonocardiogram Classification
}

\author{Martin Sondermann$^{1,2}$, Pinar Bisgin$^{1,3}$, Niklas Tschorn$^{1,4}$, Anja Burmann$^{1}$, \\ and Christoph M. Friedrich$^{2,5}$, \emph{Member,~IEEE} %
\thanks{$^{1}$Fraunhofer Institute for Software and Systems Engineering ISST, Dortmund, Germany. \{martin.sondermann, pinar.bisgin, niklas.tschorn, anja.burmann\}@isst.fraunhofer.de}%
\thanks{$^{2}$Department of Computer Science, University of Applied Sciences and Arts Dortmund, Dortmund, Germany. \mbox{christoph.friedrich@fh-dortmund.de}}%
\thanks{$^{3}$Department of Computer Science, TU Dortmund University, Dortmund, Germany.}%
\thanks{$^{4}$Department of Health Informatics, School of Medicine, Witten/Herdecke University, Witten, Germany.}%
\thanks{$^{5}$Institute for Medical Informatics, Biometry and Epidemiology (IMIBE), University Hospital Essen, Essen, Germany.}%
}

\hypersetup{
    hidelinks=true,
}
\begin{document}

\maketitle
\thispagestyle{empty}
\pagestyle{empty}

\begin{abstract}
The automated classification of phonocardiogram (PCG) recordings represents a substantial advancement in cardiovascular diagnostics.
This paper presents a systematic comparison of four distinct models for heart murmur detection: two specialized convolutional neural networks (CNNs) and two zero-shot universal audio transformers (BEATs), evaluated using fixed-length and heart cycle normalization approaches.
Utilizing the PhysioNet2022 dataset, a custom heart cycle normalization method tailored to individual cardiac rhythms is introduced.
The findings indicate the following AUROC values: the CNN model with fixed-length windowing achieves 79.5~\%, the CNN model with heart cycle normalization scores 75.4~\%, the BEATs transformer with fixed-length windowing achieves 65.7~\%, and the BEATs transformer with heart cycle normalization results in 70.1~\%.

The findings indicate that physiological signal constraints, especially those introduced by different normalization strategies, have a substantial impact on model performance.
The research provides evidence-based guidelines for architecture selection in clinical settings, emphasizing the need for a balance between accuracy and computational efficiency. 
Although specialized CNNs demonstrate superior performance overall, the zero-shot transformer models may offer promising efficiency advantages during development, such as faster training and evaluation cycles, despite their lower classification accuracy.
These findings highlight the potential of automated classification systems to enhance cardiac diagnostics and improve patient care.
\\
Index Terms — CNN, audio classification, heart sound analysis, zero-shot learning, medical signal processing, heart murmur detection, BEATs transformer, heart cycle normalization.
\end{abstract}
\vspace{0.2em}
\begin{center}
\small\textit{Preprint Version. Accepted for publication at EMBC 2025. Final version will appear in IEEE Xplore.}
\end{center}
\vspace{0.3em}
\section{Introduction}
Cardiovascular diseases are one of the leading causes of death worldwide.
In particular, untreated valvular diseases can greatly impair quality of life and lead to severe complications \cite{oliveira2021circor}.
Many cases of valvular disorders remain undetected and can contribute to premature death \cite{otto2009valvular}.
The importance of early detection is emphasized, as timely intervention can positively influence the disease course.
Heart sound analysis using phonocardiogram (PCG) recordings is a critical diagnostic tool in cardiovascular medicine, particularly for the early detection of valvular diseases and identification of murmurs.
This study systematically compares four different models for PCG classification: a specialized convolutional neural network (CNN) with fixed-length windowing and with heart cycle normalization, a BEATs transformer \cite{chen2022beats} with fixed-length windowing, and with heart cycle normalization.
By evaluating these models, the strengths and weaknesses of each approach in the context of automated cardiac diagnostics are identified, contributing to improved detection and treatment of heart conditions.
Automated classification of heart sounds presents unique challenges due to their complex temporal structure, physiological variability, and the necessity for high diagnostic accuracy in clinical settings.
Traditional diagnostics using a stethoscope require years of clinical experience for precise interpretation of heart sounds, often leading to subjective assessments \cite{krittanawong2019deep}.

While CNNs have emerged as the predominant method for PCG analysis \cite{thalmayer2020robust}, recent advancements in transformer architectures for audio processing suggest promising alternative pathways for medical signal classification \cite{gong2021ast}.
Identifying pathological heart sounds, particularly murmurs, remains a challenging task, even for experienced clinicians, with accuracy rates varying based on their expertise \cite{oliveira2021circor}.

This study aims to contribute to the development of more accurate and efficient automated classification systems that can be deployed in underserved regions.
However, several key challenges must be addressed for successful implementation. 
Variability in signal quality poses a significant barrier; the quality of PCG recordings can differ considerably due to recording equipment and environmental conditions, resulting in inconsistent and noisy signals that complicate analysis \cite{guo2001virtual}. Additionally, the diversity of patient populations introduces further challenges, as a wide range of physiological characteristics can influence the acoustic patterns captured in PCG recordings, complicating the generalization of classification models \cite{oliveira2021circor}. 
There is also a pressing need for robust performance across various cardiac pathologies, as accurate classification of different types of heart sounds, particularly murmurs, is crucial for effective diagnosis and treatment. Lastly, resource constraints in many clinical settings, especially in underserved regions, limit the computational resources and infrastructure necessary to support advanced classification systems, underscoring the need for efficient models that require minimal computational overhead \cite{fdaArtificialIntelligence2024}.

A central focus of this investigation is the implementation of heart cycle normalization as a signal preprocessing approach--a topic that has not been extensively explored in the existing literature.

\section{Related Work}
In the field of PCG analysis, both CNNs and transformer models have been explored, each offering distinct advantages and limitations.
CNNs have been widely adopted for PCG classification due to their ability to extract local features from spectrogram representations of heart sounds \cite{khanDeepLearning2021}. 
For instance, \mbox{Thalmayer et al. \cite{thalmayer2020robust}} utilized CNNs to classify heart sounds based on STFT spectrograms, demonstrating robust performance across varying physiological conditions. Similarly, \mbox{Maity et al. \cite{maity2023transfer}} explored CNNs for heart valve disease classification, leveraging transfer learning with pretrained embedding extractors for audio classification on mel spectrograms.
However, these CNN-based approaches often require extensive training data and may struggle to capture long-range temporal dependencies within the PCG signal.

Transformer models, with their self-attention characterismus \cite{vaswani2017attention}, provide an alternative approach for capturing global relationships in audio data. 
\mbox{Alkhodari et al. \cite{alkhodari2024identification}} demonstrated the effectiveness of transformer-based models for identifying congenital valvular murmurs, highlighting their ability to capture temporal dependencies over arbitrary distances. 
Nonetheless, transformers typically require the transformation of continuous audio signals into discrete tokens, which can introduce complexities and potential information loss.
\mbox{Gong et al. \cite{gong2021ast}} introduced the Audio Spectrogram Transformer (AST), showcasing the potential of transformers for various audio classification tasks. 
The authors have also explored the combination of CNNs and transformers to leverage their complementary strengths.

A key aspect that remains relatively unexplored in the existing literature is the impact of heart cycle normalization on PCG classification performance. 
While previous studies have primarily focused on fixed-length windowing or feature engineering techniques, the potential benefits of aligning PCG signals based on physiological heart cycle boundaries have not been comprehensively investigated. 
Unlike fixed-length windowing approaches, heart cycle normalization aligns the PCG signal according to physiological boundaries, which may preserve critical temporal dynamics that are otherwise lost.
This work addresses this gap by systematically comparing CNN and transformer-based models with and without heart cycle normalization, providing valuable insights into the effectiveness of this preprocessing technique for automated cardiac diagnostics. 

\section{Materials and Methods}
\subsection{Dataset and Preprocessing}
The PhysioNet2022 challenge dataset (CirCor DigiScope)\cite{oliveira2021circor} consists of $3163$ PCG recordings from $816$ unique patients.
Within the dataset, there are $156$ samples labeled as \textit{unknown}, which are excluded from this analysis.
The remaining samples are categorized as murmur positive and murmur negative.
After removing the unknown samples, $616$ $(20.49\%)$ positive and $2391$ $(79.51\%)$ negative samples are available.
Each patient may have multiple recordings, primarily from different auscultation points, resulting in a total of $816$ unique patients.
The age distribution reflects the pediatric focus of the screening campaign, with children ($2290$ recordings, $76.2\%$) representing the majority of cases, followed by adolescents ($250$, $8.3\%$), infants ($222$, $7.4\%$), and neonates ($8$, $0.27\%$). 
$737$ recordings received no age label.
The dataset shows a balanced gender distribution with $1523$ female and $1484$ male recordings.

Additionally, each recording is annotated with heart cycle phase information. In total, $62636$ complete S1 to S1 cycles are labeled, averaging $20.83$ cycles per sample.
It is important to note that not every heart cycle is labeled; only about half of the overall signal length contains annotated cycle information.
All inputs undergo a preprocessing pipeline \cite{fuadah2022optimal, singh2020short, thakur2018filtering}:
\begin{enumerate}
    \item Splitting into chunks using either the fixed or cycle method.
    \item Butterworth bandpass filtering ($25~Hz \leq f \leq 500~Hz$, $5$th order).
    \item Min-max amplitude normalization to the range $[0,1]$.
    \item Signal augmentation with $60\%$ probability for each method:
    \begin{itemize}
        \item Random muting of up to $25\%$ of the signal length
        \item Pitch shifting by up to $\pm1$ semitone while maintaining physiological boundaries
        \item For CNN inputs: Random masking of up to $25\%$ of the spectrogram area
    \end{itemize}
\end{enumerate}
The fixed chunk method splits the signal based on a predefined duration.
Every \textit{seconds} $\times$ \textit{sample\_rate} samples create a new chunk.
If at the end the remaining chunk is shorter than the configured \textit{seconds} parameter but contains more than $65\%$ of the samples of a full chunk, it is padded with the median value of the current chunk.
Otherwise, it is discarded.

\subsection{Heart Cycle based Normalization}
The cycle normalization process transforms variable-length PCG recordings into uniform segments while preserving physiological heart cycle boundaries.
This approach differs fundamentally from fixed-duration windowing by utilizing annotated S1 peak timestamps as natural segment delimiters.
Each normalized chunk contains a predefined number of complete heart cycles, which are collectively stretched or compressed to achieve a consistent sample count.

The normalization process identifies complete heart cycles between consecutive S1 peaks in the signal. A sequence of N consecutive cycles ($N=10$ or $N=12$ depending on configuration) forms a single chunk. If insufficient cycles are available, the chunk is discarded rather than partially filled. 
The identified sequence undergoes time-domain stretching to achieve a uniform sample count across all chunks, determined by the \textit{seconds} and \textit{sample\_rate} parameters, enabling consistent model input dimensions

\begin{figure}[h]
\centering
\includegraphics[width=\columnwidth]{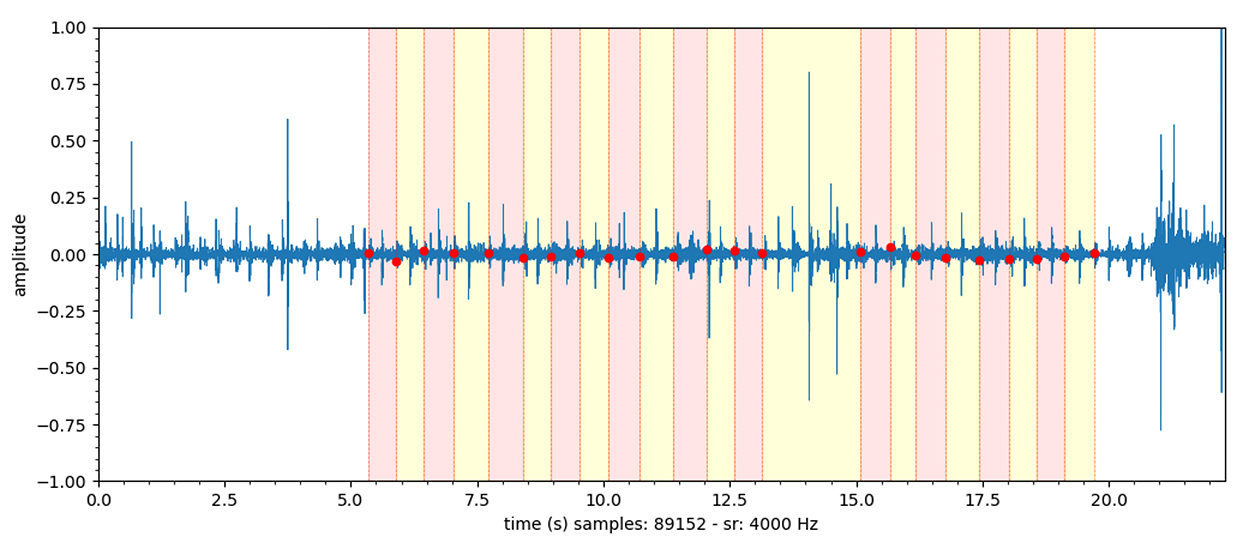}
\caption{Example signal after applying filtering and normalization. Red dots show the annotated S1 peaks. Alternating red and yellow areas represent the partitions of the signal between different heart cycles. This example illustrates the incomplete annotation and incorrectly long cycle between beat \mbox{$14$ and $15$.}}
\label{fig:cycle_stretch_problem}
\end{figure}

A considerable challenge in this approach emerges from inconsistent S1 peak annotations.
Missing annotations create artificially extended cycles, as demonstrated in Fig.~\ref{fig:cycle_stretch_problem}.
Red dots illustrate the annotated S1 peaks and alternating red-yellow areas display the area between each cycle.
The area between cycle $14$ and $15$ is visibly larger because of a missing cycle annotation.
These gaps result in non-physiological cycle durations that can distort the signal during the stretching process.
In this implementation no validation or checks to detect such cases are applied.
Chunks containing fewer cycles than required are simply discarded.

The time-domain stretching operation utilizes librosa's time-stretch function\footnote{Available at: \url{https://librosa.org/doc/main/generated/librosa.effects.time_stretch.html} (Last accessed: February 3, 2025).}, which modifies the temporal duration of an audio signal while preserving critical spectral characteristics. The stretch factor for each chunk is computed as follows:

\begin{equation}
    \text{stretch\_factor} = \frac{\text{target\_duration}}{\text{cycle\_duration}} \cdot \frac{\text{target\_sr}}{\text{original\_sr}}
\end{equation}

\noindent In this equation, \textit{stretch\_factor} represents the ratio by which the audio signal's duration will be modified.
The variable~\textit{target\_duration} indicates the desired duration that the audio signal should achieve after the stretching process.
Conversely,~\textit{cycle\_duration} refers to the original time span of $N$ consecutive cycles in the audio signal, which serves as a reference for the stretching operation.
\textit{\text{target\_sr}} and \textit{\text{original\_sr}} notate the corresponsding samplerate.

\noindent Additionally, \textit{target\_sr} denotes the desired sampling rate that the audio signal should be converted to, while \textit{original\_sr} represents the original sampling rate of the audio signal.
By calculating the stretch factor using these variables, the time-domain stretching operation effectively alters the duration of the audio while maintaining its essential spectral features.
This approach achieves physiological alignment at the cost of data utilization efficiency - only $51.7\%$ of the audio samples contain sufficient annotated cycles for processing.
This reduction is not a feature of the normalization method but a necessary consequence of the incomplete annotations in the dataset.
The here used method does not actively identify and remove unreliable data, but rather requires complete cycle annotations to function properly.
While this represents a substantial reduction in available training data, it ensures that all processed segments maintain consistent physiological structure and temporal alignment.

\section{Model Architecture}
Two primary model architectures for the automated classification of PCG recordings have been developed: a CNN and a BEATs transformer model \cite{chen2022beats}.
\subsection{CNN}
The CNN implementation utilizes a deep architecture specifically designed for PCG signal processing, comprising four sequential convolutional blocks followed by a classification head.
Each convolutional block employs $3\times3$ kernels with consistent stride and padding configurations, progressively increasing feature map depths from $8$ to $128$ channels while reducing spatial dimensions through max-pooling operations.
All parameters and network shapes are chosen empirically after consulting relevant literature \cite{tsalera_comparison_2021,maity2023transfer,singh2020short,thalmayer2020robust}.
\begin{table}[ht]
\caption{CNN model specifications.}
\label{tab:cnn_specs}
\centering
\footnotesize 
\setlength{\tabcolsep}{0.3em}
\begin{tabular}{|l|p{2.5cm}|p{2.5cm}|}
\hline
\textbf{Parameter} & \textbf{Fixed Window} & \textbf{Cycle Normalized} \\
\hline
Input format & $8.0~s$ & $10$ cycles \\
Mel bands & $352$ & $128$ \\
FFT size & $512$ & $1152$ \\
Hop length & $352$ & $288$ \\
\hline
\multicolumn{3}{|c|}{\textbf{Common Architecture}} \\
\hline
Conv. channels & \multicolumn{2}{c|}{$1$→$8$→$16$→$64$→$128$} \\
Kernel size & \multicolumn{2}{c|}{$(3,3)$} \\
Stride & \multicolumn{2}{c|}{$(1,1)$} \\
Padding & \multicolumn{2}{c|}{$(0,0)$} \\
Max-pooling & \multicolumn{2}{c|}{$(2,2)$ after each conv block} \\
Adaptive pooling output & \multicolumn{2}{c|}{$(4,4)$} \\
FC layers & \multicolumn{2}{c|}{$2048$→$512$→$2$ with dropout $(0.4, 0.6)$} \\
\hline
\end{tabular}
\end{table}
The mel band resolution offers a compromise between frequency and time resolution as well as computational efficiency.
Both configurations share identical convolutional architectures but differ in their front-end signal processing parameters.
The SiLU activation function was chosen for its ability to mitigate the \mbox{'dying ReLU'} problem and provide smooth non-linearities, which can enhance learning performance in deep networks\cite{huang_deriving_2025}.
It is followed by max-pooling for spatial reduction.
The classification head employs adaptive average pooling to ensure consistent feature dimensions, followed by three fully connected layers with batch normalization and dropout regularization.
This architecture balances feature extraction capacity with computational efficiency, while maintaining robust generalization through systematic regularization techniques \cite{ioffe_batch_2015}.

\subsection{BEATs}
The BEATs transformer \cite{chen2022beats} approach utilizes a zero-shot audio classification pipeline, combining a pre-trained audio transformer for embedding extraction with a k-Nearest-Neighbor (k-NN) classifier.
The implementation leverages the publicly available \textit{BEATs-iter3+-AS2M} weights, which were pre-trained by Chen et al. on the AudioSet dataset \cite{gemmekeAudioSet2017}.
These pre-trained weights were used directly to generate fixed-dimensional embeddings without any additional fine-tuning on the PCG data.
\begin{table}[h]
\caption{BEATs-kNN configuration details.}
\label{tab:beats_specs}
\centering
\footnotesize 
\setlength{\tabcolsep}{0.3em} 
\begin{tabular}{|l|p{2.5cm}|p{2.5cm}|} 
\hline
\textbf{Parameter} & \textbf{Fixed Window} & \textbf{Cycle Normalized} \\
\hline
Input format & $7.0~s$ & $12$ cycles \\
Extractor type & BEATs-iter3+-AS2M & BEATs-iter3+-AS2M \\
Embedding dim & $768$ & $768$ \\
Target sample rate & $16~kHz$ & $16~kHz$ \\
\hline
\multicolumn{3}{|c|}{\textbf{Classifier Configuration}} \\
\hline
\multicolumn{3}{|c|}{k-NN neighbors: $5$ (fixed), $7$ (cycle)} \\
\multicolumn{3}{|c|}{Distance metric: Euclidean} \\
\multicolumn{3}{|c|}{Weight function: Uniform} \\
\hline
\end{tabular}
\end{table}

The implementation, detailed in Tab. \ref{tab:beats_specs}, processes audio inputs in two stages.
First, the BEATs transformer extracts $768$-dimensional embeddings from the input signals, maintaining a consistent representation regardless of input length.
These embeddings are stored for the entire training dataset before classification begins.
The second stage employs a k-NN classifier operating directly in the embedding space, with k values optimized separately for fixed and heart cycle-normalized approaches.

Initial experiments explored dimensionality reduction through UMAP \cite{mcinnesUMAPUniform2018} and class balancing via SMOTE \cite{chawlaSMOTESynthetic2002} before classification.
However, these preprocessing steps did not yield consistent improvements in classification performance and were omitted from the final implementation.
The k-NN classifier operates directly on the raw embeddings, using Euclidean distance with uniform weighting for neighbor voting, providing a simple yet effective classification approach that leverages the pre-trained audio representations.

\subsection{Architecture Training}
The evaluation methodology employs a comprehensive 10-fold cross-validation (CV) strategy with patient-wise stratification and grouping to ensure robust performance assessment.
This approach prevents data leakage by maintaining complete patient separation between training and validation sets.

\noindent \textbf{Handling Class Imbalance.} The class imbalance ($20.49\%$ positive cases) is addressed through different strategies: CNN training uses the more robust FocalLoss criterion \cite{lin_focal_2020} with the class balance set as the $\alpha$ parameter and $\gamma=2$.
For the BEATs-kNN approach, this is not feasible, so the optimisation focused on setting the $k$-Parameter and experimenting with different implementations in combination with UMAP and SMOTE.
All evaluations uphold stratified fold distributions to ensure consistent class representation across training and validation datasets.

The CNN models utilize different training configurations optimized for their respective windowing approaches.
Training parameters are detailed in Tab. \ref{tab:train_params}.
\begin{table}[h]
\caption{Training parameters for the CNN models.}
\label{tab:train_params}
\centering
\footnotesize
\setlength{\tabcolsep}{0.3em}
\begin{tabular}{|l|p{2.5cm}|p{2.5cm}|}
\hline
\textbf{Parameter} & \textbf{Fixed window} & \textbf{Cycle normalized} \\
\hline
Optimizer & AdamW & Adam \\
Learning rate & $1\text{e-}4$ & $1\text{e-}3$ \\
Batch size & $32$ & $32$ \\
Scheduler & Reduce on plateau & StepLR \\
Scheduler patience & $10$ & $10$ \\
Scheduler factor & $0.5$ & $0.5$ \\
L1 regularization & $1\text{e-}5$ & $0.0$ \\
L2 regularization & $5\text{e-}4$ & $0.003$ \\
Epochs & $60$ & $60$ \\
\hline
\end{tabular}
\end{table}
The BEATs-kNN approach requires minimal training configuration as it utilizes frozen pre-trained weights. 
Due to the simple inferencing method, the training is finished after one epoch.
\subsection{Evaluation metrics}
A comprehensive set of evaluation metrics suitable for imbalanced medical classification tasks was employed to assess the performance of the models developed in this study. These metrics offer complementary perspectives, focusing on clinical relevance in heart sound classification. The following core metrics were utilized for binary classification of heart sounds:\\
\textit{Precision:} Measures the proportion of true positive predictions among all positive predictions, indicating the model's effectiveness in minimizing false positives \cite{s21196656}.\\
\textit{Recall:} Also known as sensitivity, this metric assesses the model's ability to identify all relevant true positives, which is crucial in medical diagnostics \cite{s21196656}.\\
\textit{Area Under the Receiver Operating Characteristic curve (AUROC):} Evaluates the model's discrimination ability across all prediction thresholds. An AUROC value closer to 1 indicates better separation between positive and negative classes.\\
\textit{Matthews Correlation Coefficient (MCC):} Provides a balanced measure for imbalanced datasets, ranging from $-1$ (total disagreement) to $+1$ (perfect prediction). An MCC value of $0$ represents random guessing \cite{chiccoMatthewsCorrelation2023}.\\
\textit{F2-Score:} This metric gives additional weight to recall, which is critical in medical diagnostics, especially when missing positive cases has significant costs \cite{chiccoMatthewsCorrelation2023}:
\begin{equation}\label{eq:f2_score} \text{F2-Score} = 5 \cdot \frac{\text{Precision} \cdot \text{Recall}}{4 \cdot \text{Precision} + \text{Recall}} \end{equation}

The evaluation of these metrics allows for a comprehensive assessment of each model's performance in classifying heart sounds, highlighting the strengths and weaknesses of the different architectures used.
\section{Results}
\subsection{Architecture Results}
\begin{figure*}[p]
\centering
\begin{subfigure}{0.45\textwidth}
  \centering
  \includegraphics[width=0.7\textwidth]{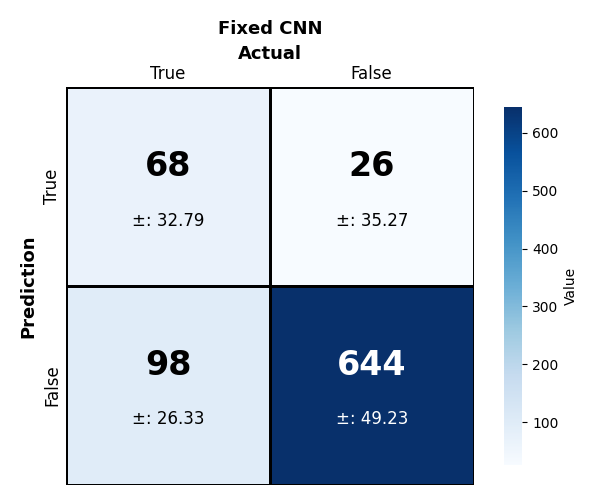}
  \caption{}
  \label{fig:subfig1}
\end{subfigure}%
\begin{subfigure}{0.45\textwidth}
  \centering
  \includegraphics[width=0.7\textwidth]{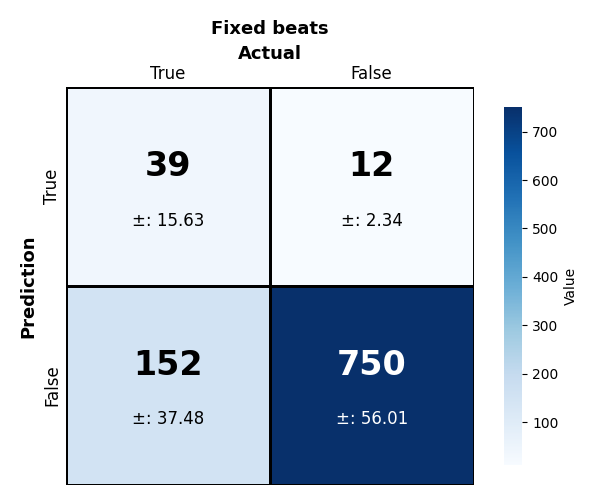}
  \caption{}
  \label{fig:subfig2}
\end{subfigure}
\begin{subfigure}{0.45\textwidth}
  \centering
  \includegraphics[width=0.7\textwidth]{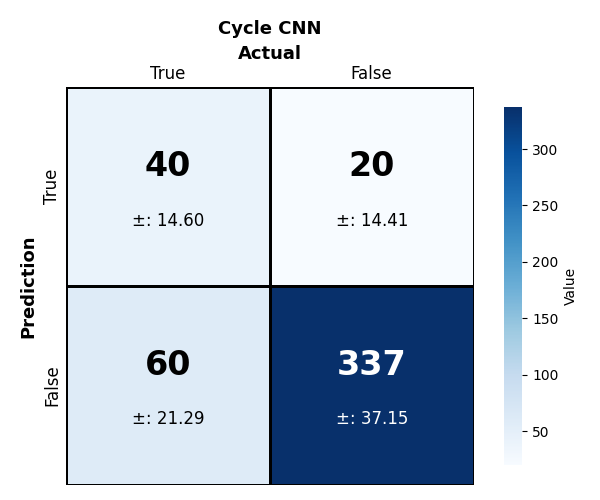}
  \caption{}
  \label{fig:subfig3}
\end{subfigure}%
\begin{subfigure}{0.45\textwidth}
  \centering
  \includegraphics[width=0.7\textwidth]{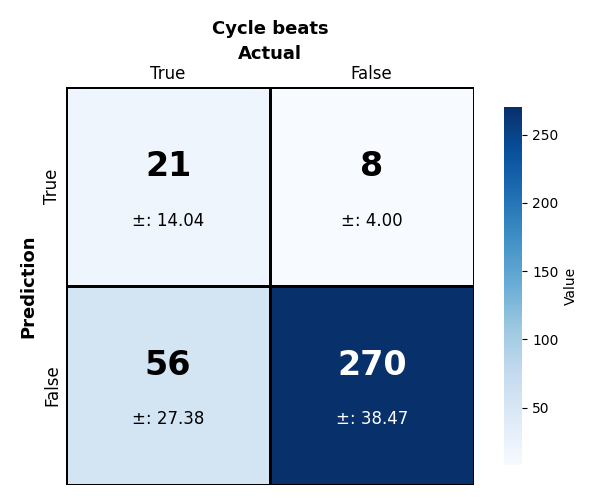}
  \caption{}
  \label{fig:subfig4}
\end{subfigure}
\caption{Validation confusion matrices for CNN and BEATs models, for fixed and cycle window methods: (a)~CNN with fixed-length window, (b)~BEATs with fixed-length window, (c)~CNN with cycle normalization, and (d)~BEATs with cycle normalization. CNN models shown at epoch 60, BEATs models after single epoch extraction.
All values represent means across 10-fold cross-validation with their standard deviations.}
\label{fig:cms}
\end{figure*}

For the classification of heart sounds in the PhysioNet2022 dataset, various models were evaluated, including CNN and BEATs-kNN architectures. The confusion matrices presented in Fig. \ref{fig:cms} illustrate the distribution of classification results for these models.
Key observations from the matrices reveal that the CNN model demonstrates a significant number of true positive (TP) predictions, indicating effective identification of heart murmurs.
Conversely, the BEATs model displays a higher rate of false negative (FN) predictions, suggesting challenges in recognizing certain positive cases. Additionally, varying preprocessing techniques, such as window size adjustments, yield differing results across models.
For instance, the CNN's performance slightly decreases with cycle-normalized window size, while the BEATs model improves in its ability to identify positive cases under the same conditions.

The key parameters and results of the classifiers are summarized in Tab. \ref{tab:performance}. The results indicate that the CNN model with fixed-length window size achieves an AUROC of $0.795$, with a precision of $0.811$ and a recall of $0.392$. These values suggest a robust performance of the model. The CNN model with cycle-normalized window size achieves an AUROC of $0.754$, with a precision of $0.690$ and a recall of $0.411$. For the BEATs-kNN model, the fixed-length window size variant achieves an AUROC of $0.657$, with a precision of $0.731$ and a recall of $0.202$. Meanwhile, the BEATs-kNN model with cycle-normalized window size achieves an AUROC of $0.701$, with a precision of $0.717$ and a recall of $0.272$.

\begin{table*}[p]
\centering
\caption{Performance comparison across architectures. Average and std. of 10-fold CV.}
\label{tab:performance}
\setlength{\tabcolsep}{0.5em} 
\large
\begin{tabular}{|l|c|c|c|c|c|}
\hline
\textbf{Architecture} & \textbf{AUROC} & \textbf{MCC} & \textbf{Precision} & \textbf{Recall} & \textbf{F2} \\
\hline
CNN fixed & $0.795 \pm 0.049$ & $0.472 \pm 0.090$ & $0.811 \pm 0.169$ & $0.392 \pm 0.173$ & $0.437 \pm 0.157$ \\
CNN cycle & $0.754 \pm 0.077$ & $0.432 \pm 0.122$ & $0.690 \pm 0.161$ & $0.411 \pm 0.120$ & $0.447 \pm 0.112$ \\
BEATs fixed & $0.657 \pm 0.053$ & $0.322 \pm 0.096$ & $0.731 \pm 0.118$ & $0.202 \pm 0.077$ & $0.235 \pm 0.086$ \\
BEATs cycle & $0.701 \pm 0.045$ & $0.362 \pm 0.084$ & $0.717 \pm 0.131$ & $0.272 \pm 0.074$ & $0.311 \pm 0.080$ \\
\hline
\end{tabular}
\end{table*}

The CNN with fixed-length windowing demonstrates superior discrimination ability (AUROC $= 0.795 \pm 0.049$) compared to other approaches.
Notably, cycle normalization shows divergent effects between architectures: while slightly degrading CNN performance, it improves BEATs model metrics.

\subsection{Impact of Heart Cycle Normalization}
The effect of heart cycle normalization demonstrates notably different impacts across architectures.
The specialized CNN architecture experiences a decrease in discriminative power, with AUROC reducing from $0.795$ to $0.754$ under cycle normalization.
Similarly, the MCC drops \mbox{from $0.472$ to $0.432$}, indicating a slight degradation in overall classification performance.
The fixed-length windowing approach may have provided a more consistent input for the CNN to learn from, compared to the variable-length cycles used in the cycle normalization approach and time-domain stretching introduced artifacts that disrupted the CNN's feature extraction process.
 
\begin{figure*}[p]
\centering
\includegraphics[width=0.95\textwidth]{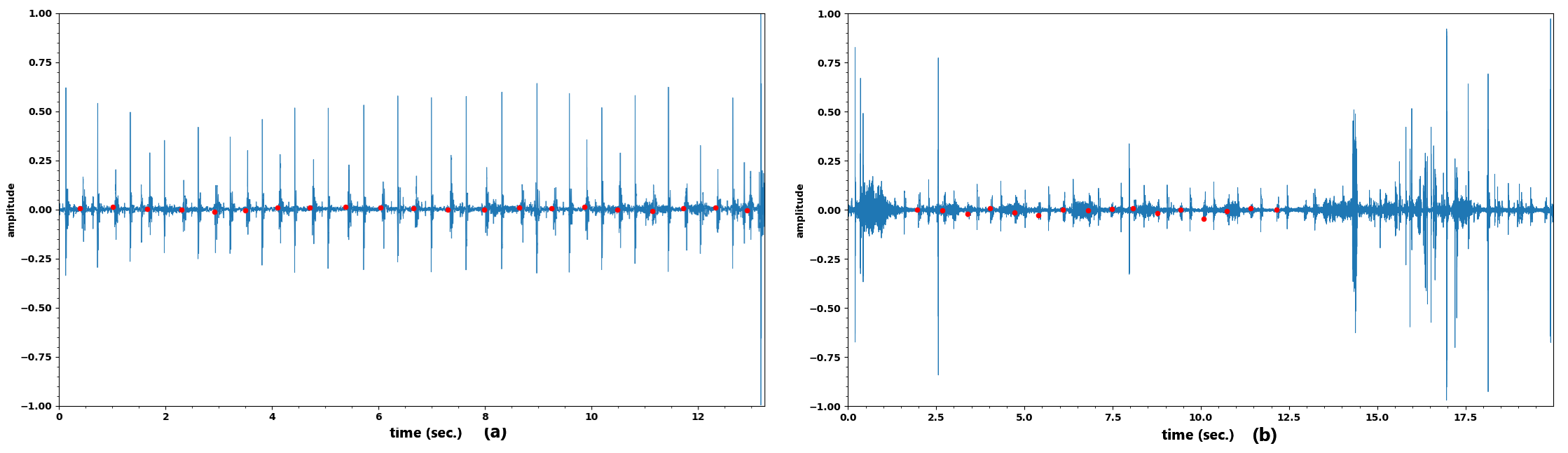}
\caption{Comparison of PCG recordings with varying annotation quality: Red dots indicate annotated S1 peaks. (a) ideal case with a consistently annotated signal and clear distinguishable heart cycles, (b) partially annotated signal showing missing chunks without any annotation and noise perturbations.}\label{fig:cycle_analysis}
\end{figure*}

Conversely, the BEATs-kNN approach shows improved performance with cycle normalization, increasing AUROC from $0.657$ to $0.701$ and MCC from $0.322$ to $0.362$.
This divergent behavior suggests that the pre-trained audio representations might better capture physiological patterns when presented with cycle-normalized inputs.
Fig. \ref{fig:cycle_analysis} presents a comparison of PCG recordings with varying annotation quality: Red dots indicate annotated S1 peaks.
(a) shows a consistently annotated signal with clear, distinguishable heart cycles, (b) shows a signal with noisy segments and incomplete annotations, resulting in missing S1 peaks and less clear heart cycle boundaries.

These annotation inconsistencies, including the one mentioned in Fig. \ref{fig:cycle_stretch_problem}, directly impact data utilization - segments with insufficient annotated cycles must be discarded to maintain physiological validity.
This constraint particularly affects the CNN model's performance, which benefits from larger training datasets.
However, the BEATs model's improved performance under cycle normalization suggests that maintaining physiological cycle boundaries might enable better feature extraction from the pre-trained audio representations, despite the overall reduced data availability.

\section{Discussion}
The performance differences between CNN and BEATs architectures highlight several key considerations in the context of heart sound classification.
CNNs excel at learning task-specific features; however, they require extensive training on labeled datasets to achieve optimal performance.
On the other hand, the BEATs transformer model offers a zero-shot classification capability, which markedly reduces computational overhead during training and thus allows for greater flexibility in development.
Additionally, the benefits of cycle normalization vary based on the architectural approach employed, with CNNs potentially enhancing their ability to capture physiological characteristics through normalization.

Despite these strengths, this study reveals several limitations that provide opportunities for future research. 
A primary challenge stems from the quality of PCG annotations in the PhysioNet2022 dataset, where only 51.7\% of recordings contain complete heart cycle annotations.
This limitation constrains the effectiveness of physiological signal normalization and may mask its full benefits.
Variability in recording quality across different clinical sites presents another considerable challenge, leading to inconsistencies in background noise levels and signal characteristics.
While this preprocessing pipeline includes standardized filtering and normalization steps, these fundamental differences may still influence model performance.

The dataset's pathological diversity represents a third limitation, as the binary classification approach simplifies the complex spectrum of cardiac pathologies.
Future work should explore multi-class classification methods to better differentiate between specific heart conditions, requiring additional labeled data with detailed pathological annotations.

These limitations suggest several promising research directions, including the development of robust automated cycle detection methods to reduce dependence on manual annotations, the integration of signal quality assessment metrics to weight samples during training, and the investigation of few-shot learning approaches to address rare pathological conditions.
Furthermore, exploring self-supervised pre-training specifically for cardiac audio could enhance model performance.
The potential of ensemble methods that combine both CNN and transformer-based approaches should also be investigated, leveraging their complementary strengths. 

\section{Conclusion}
This study demonstrates that while specialized CNNs currently achieve superior performance in PCG classification, zero-shot transformer models offer a promising alternative, particularly in resource-constrained settings.
The custom heart cycle normalization approach improved the performance of the transformer model, highlighting the importance of considering physiological signal characteristics in medical audio analysis and selecting the appropriate architecture for a given context of application.
By systematically comparing these four models, valuable insights are provided to guide future developments in automated cardiac diagnostics, particularly in the development of efficient and accurate diagnostic tools.
The limitations identified in this work provide a roadmap for future research, particularly toward developing robust annotation methods and optimizing physiological signal normalization techniques for cardiac audio analysis.
\vspace{1.3em}
\\
\noindent\textsf{\textit{Author Statement}}
\vspace{0.4em}
\\
This study exclusively used the publicly available \textit{CirCor DigiScope Phonocardiogram Dataset}, which was previously collected with appropriate ethical approval.
\bibliographystyle{IEEEtran}
\bibliography{ieee_ma}

\begin{thebibliography}{10}
\providecommand{\url}[1]{#1}
\csname url@samestyle\endcsname
\providecommand{\newblock}{\relax}
\providecommand{\bibinfo}[2]{#2}
\providecommand{\BIBentrySTDinterwordspacing}{\spaceskip=0pt\relax}
\providecommand{\BIBentryALTinterwordstretchfactor}{4}
\providecommand{\BIBentryALTinterwordspacing}{\spaceskip=\fontdimen2\font plus
\BIBentryALTinterwordstretchfactor\fontdimen3\font minus
  \fontdimen4\font\relax}
\providecommand{\BIBforeignlanguage}[2]{{%
\expandafter\ifx\csname l@#1\endcsname\relax
\typeout{** WARNING: IEEEtran.bst: No hyphenation pattern has been}%
\typeout{** loaded for the language `#1'. Using the pattern for}%
\typeout{** the default language instead.}%
\else
\language=\csname l@#1\endcsname
\fi
#2}}
\providecommand{\BIBdecl}{\relax}
\BIBdecl

\bibitem{oliveira2021circor}
J.~Oliveira, F.~Renna, P.~D. Costa, M.~Nogueira, C.~Oliveira, C.~Ferreira,
  A.~Jorge, S.~Mattos, T.~Hatem, T.~Tavares, A.~Elola, A.~B. Rad, R.~Sameni,
  G.~D. Clifford, and M.~T. Coimbra, ``The {{CirCor DigiScope Dataset}}: {{From
  Murmur Detection}} to {{Murmur Classification}},'' \emph{IEEE journal of
  biomedical and health informatics}, vol.~26, no.~6, pp. 2524--2535, Jun.
  2022, doi: 10.1109/JBHI.2021.3137048.

\bibitem{otto2009valvular}
C.~M. Otto and R.~O. Bonow, ``Valvular heart disease : A companion to
  {{Braunwald}}'s heart disease,'' 2020, iSBN: 978-0-323-54633-1 , published at
  Elsevier, 5th Edition.

\bibitem{chen2022beats}
\BIBentryALTinterwordspacing
S.~Chen, Y.~Wu, C.~Wang, S.~Liu, D.~Tompkins, Z.~Chen, W.~Che, X.~Yu, and
  F.~Wei, ``{BEATs}: Audio pre-training with acoustic tokenizers,'' in
  \emph{Proceedings of the 40th International Conference on Machine Learning},
  vol. 202.\hskip 1em plus 0.5em minus 0.4em\relax {PMLR}, 2023, pp.
  5178--5193. [Online]. Available:
  \url{https://proceedings.mlr.press/v202/chen23ag.html}
\BIBentrySTDinterwordspacing

\bibitem{krittanawong2019deep}
\BIBentryALTinterwordspacing
C.~Krittanawong, K.~W. Johnson, R.~S. Rosenson, Z.~Wang, M.~Aydar, U.~Baber,
  J.~K. Min, W.~H.~W. Tang, J.~L. Halperin, and S.~M. Narayan, ``Deep learning
  for cardiovascular medicine: A practical primer,'' \emph{European Heart
  Journal}, vol.~40, no.~25, pp. 2058--2073, Jul. 2019, doi:
  10.1093/eurheartj/ehz056. [Online]. Available:
  \url{https://www.ncbi.nlm.nih.gov/pmc/articles/PMC6600129/}
\BIBentrySTDinterwordspacing

\bibitem{thalmayer2020robust}
\BIBentryALTinterwordspacing
A.~Thalmayer, S.~Zeising, G.~Fischer, and J.~Kirchner, ``A {{Robust}} and
  {{Real-Time Capable Envelope-Based Algorithm}} for {{Heart Sound
  Classification}}: {{Validation}} under {{Different Physiological
  Conditions}},'' \emph{Sensors (Basel, Switzerland)}, vol.~20, no.~4, p. 972,
  Feb. 2020, doi: 10.3390/s20040972. [Online]. Available:
  \url{https://www.ncbi.nlm.nih.gov/pmc/articles/PMC7070375/}
\BIBentrySTDinterwordspacing

\bibitem{gong2021ast}
Y.~Gong, Y.-A. Chung, and J.~Glass, ``{{AST}}: {{Audio Spectrogram
  Transformer}},'' \emph{Proc. {{Interspeech}} 2021}, pp. 571--575, Jul. 2021,
  brno, Czech Republic. doi: 10.21437/Interspeech.2021-698.

\bibitem{guo2001virtual}
Z.~Guo, C.~Moulder, Y.~Zou, M.~Loew, and L.~G. Durand, ``A virtual instrument
  for acquisition and analysis of the phonocardiogram and its internet-based
  application,'' \emph{Telemedicine Journal and E-Health}, vol.~7, no.~4, pp.
  333--339, 2001, doi: 10.1089/15305620152814737.

\bibitem{fdaArtificialIntelligence2024}
{FDA}, ``Artificial intelligence \& medical products,'' March 2024, available
  November 19, 2024 at \url{https://www.fda.gov/media/177030/download}.

\bibitem{khanDeepLearning2021}
\BIBentryALTinterwordspacing
K.~N. Khan, F.~A. Khan, A.~Abid, T.~Olmez, Z.~Dokur, A.~Khandakar, M.~E.~H.
  Chowdhury, and M.~S. Khan, ``Deep learning based classification of
  unsegmented phonocardiogram spectrograms leveraging transfer learning,''
  \emph{Physiological Measurement}, vol.~42, no.~9, pp. 1--22, Sep. 2021, doi:
  10.1088/1361-6579/ac1d59. [Online]. Available:
  \url{https://iopscience.iop.org/article/10.1088/1361-6579/ac1d59}
\BIBentrySTDinterwordspacing

\bibitem{maity2023transfer}
\BIBentryALTinterwordspacing
A.~Maity, A.~Pathak, and G.~Saha, ``Transfer learning based heart valve disease
  classification from {{Phonocardiogram}} signal,'' \emph{Biomedical Signal
  Processing and Control}, vol.~85, pp. 1--17, Aug. 2023, doi:
  10.1016/j.bspc.2023.104805. [Online]. Available:
  \url{https://www.sciencedirect.com/science/article/pii/S1746809423002380}
\BIBentrySTDinterwordspacing

\bibitem{vaswani2017attention}
\BIBentryALTinterwordspacing
A.~Vaswani, N.~Shazeer, N.~Parmar, J.~Uszkoreit, L.~Jones, A.~N. Gomez,
  L.~Kaiser, and I.~Polosukhin, ``Attention {{Is All You Need}},'' Aug. 2023.
  [Online]. Available: \url{http://arxiv.org/abs/1706.03762v7}
\BIBentrySTDinterwordspacing

\bibitem{alkhodari2024identification}
M.~Alkhodari, L.~J. Hadjileontiadis, and A.~H. Khandoker, ``Identification of
  {{Congenital Valvular Murmurs}} in {{Young Patients Using Deep Learning-Based
  Attention Transformers}} and {{Phonocardiograms}},'' \emph{IEEE Journal of
  Biomedical and Health Informatics}, vol.~28, no.~4, pp. 1803--1814, Apr.
  2024, doi: 10.1109/JBHI.2024.3357506.

\bibitem{fuadah2022optimal}
\BIBentryALTinterwordspacing
Y.~N. Fuadah, M.~A. Pramudito, and K.~M. Lim, ``An {{Optimal Approach}} for
  {{Heart Sound Classification Using Grid Search}} in {{Hyperparameter
  Optimization}} of {{Machine Learning}},'' \emph{Bioengineering}, vol.~10,
  no.~1, p.~45, Jan. 2023, doi: 10.3390/bioengineering10010045. [Online].
  Available: \url{https://www.mdpi.com/2306-5354/10/1/45}
\BIBentrySTDinterwordspacing

\bibitem{singh2020short}
\BIBentryALTinterwordspacing
S.~A. Singh, T.~G. Meitei, and S.~Majumder, ``6 - {{Short PCG}} classification
  based on deep learning,'' in \emph{Deep {{Learning Techniques}} for
  {{Biomedical}} and {{Health Informatics}}}, ser. 1, B.~Agarwal, V.~E. Balas,
  L.~C. Jain, R.~C. Poonia, and {Manisha}, Eds.\hskip 1em plus 0.5em minus
  0.4em\relax Academic Press, Jan. 2020, vol.~1, pp. 141--164, doi:
  10.1016/B978-0-12-819061-6.00006-9. [Online]. Available:
  \url{https://www.sciencedirect.com/science/article/pii/B9780128190616000069}
\BIBentrySTDinterwordspacing

\bibitem{thakur2018filtering}
\BIBentryALTinterwordspacing
R.~Thakur, M.~K. Pandey, and N.~Gupta, ``Filtering of {{Noise}} in
  {{Audio}}/{{Voice Signal}},'' in \emph{2018 3rd {{International Conference}}
  on {{Contemporary Computing}} and {{Informatics}} ({{IC3I}})}, Gurgaon,
  India, Oct. 2018, pp. 119--123, doi: 10.1109/IC3I44769.2018.9007299.
  [Online]. Available: \url{https://ieeexplore.ieee.org/document/9007299}
\BIBentrySTDinterwordspacing

\bibitem{tsalera_comparison_2021}
\BIBentryALTinterwordspacing
E.~Tsalera, A.~Papadakis, and M.~Samarakou, ``Comparison of {{Pre-Trained
  CNNs}} for {{Audio Classification Using Transfer Learning}},'' \emph{Journal
  of Sensor and Actuator Networks}, vol.~10, no.~4, pp. 1--22, Dec. 2021, doi:
  10.3390/jsan10040072. [Online]. Available:
  \url{https://www.mdpi.com/2224-2708/10/4/72}
\BIBentrySTDinterwordspacing

\bibitem{huang_deriving_2025}
\BIBentryALTinterwordspacing
A.~H. Huang and I.~Schlag, ``Deriving activation functions using integration,''
  2025. [Online]. Available: \url{http://arxiv.org/abs/2411.13010v3}
\BIBentrySTDinterwordspacing

\bibitem{ioffe_batch_2015}
\BIBentryALTinterwordspacing
S.~Ioffe and C.~Szegedy, ``Batch normalization: accelerating deep network
  training by reducing internal covariate shift,'' in \emph{Proceedings of the
  32nd International Conference on International Conference on Machine Learning
  - Volume 37}, ser. {ICML}'15.\hskip 1em plus 0.5em minus 0.4em\relax
  {JMLR}.org, pp. 448--456. [Online]. Available:
  \url{https://dl.acm.org/doi/10.5555/3045118.3045167}
\BIBentrySTDinterwordspacing

\bibitem{gemmekeAudioSet2017}
J.~F. Gemmeke, D.~P.~W. Ellis, D.~Freedman, A.~Jansen, W.~Lawrence, R.~C.
  Moore, M.~Plakal, and M.~Ritter, ``Audio {{Set}}: {{An}} ontology and
  human-labeled dataset for audio events,'' in \emph{Proc. {{IEEE ICASSP}}
  2017}.\hskip 1em plus 0.5em minus 0.4em\relax New Orleans, LA: IEEE, 2017,
  pp. 776--780, doi: 10.1109/ICASSP.2017.7952261.

\bibitem{mcinnesUMAPUniform2018}
\BIBentryALTinterwordspacing
L.~McInnes, J.~Healy, N.~Saul, and L.~Gro{\ss}berger, ``{{UMAP}}: {{Uniform
  Manifold Approximation}} and {{Projection}},'' \emph{Journal of Open Source
  Software}, vol.~3, no.~29, p. 861, Sep. 2018, doi: 10.21105/joss.00861.
  [Online]. Available: \url{https://joss.theoj.org/papers/10.21105/joss.00861}
\BIBentrySTDinterwordspacing

\bibitem{chawlaSMOTESynthetic2002}
\BIBentryALTinterwordspacing
N.~V. Chawla, K.~W. Bowyer, L.~O. Hall, and W.~P. Kegelmeyer, ``{{SMOTE}}:
  {{Synthetic Minority Over-sampling Technique}},'' \emph{Journal of Artificial
  Intelligence Research}, vol.~16, pp. 321--357, Jun. 2002, doi:
  10.1613/jair.953. [Online]. Available:
  \url{https://www.jair.org/index.php/jair/article/view/10302}
\BIBentrySTDinterwordspacing

\bibitem{lin_focal_2020}
\BIBentryALTinterwordspacing
T.-Y. Lin, P.~Goyal, R.~Girshick, K.~He, and P.~Dollár, ``Focal loss for dense
  object detection,'' \emph{{IEEE} Transactions on Pattern Analysis and Machine
  Intelligence}, vol.~42, no.~2, pp. 318--327, 2020, doi:
  10.1109/TPAMI.2018.2858826. [Online]. Available:
  \url{https://ieeexplore.ieee.org/document/8417976}
\BIBentrySTDinterwordspacing

\bibitem{s21196656}
\BIBentryALTinterwordspacing
R.~Salvi, P.~Fuentealba, J.~Henze, P.~Bisgin, T.~Sühn, M.~Spiller, A.~Burmann,
  A.~Boese, A.~Illanes, and M.~Friebe, ``Vascular auscultation of carotid
  artery: Towards biometric identification and verification of individuals,''
  \emph{Sensors}, vol.~21, no.~19, 2021, doi: 10.3390/s21196656. [Online].
  Available: \url{https://www.mdpi.com/1424-8220/21/19/6656}
\BIBentrySTDinterwordspacing

\bibitem{chiccoMatthewsCorrelation2023}
\BIBentryALTinterwordspacing
D.~Chicco and G.~Jurman, ``The {{Matthews}} correlation coefficient~({{MCC}})
  should replace the {{ROC~AUC}} as the standard metric for assessing binary
  classification,'' \emph{BioData Mining}, vol.~16, no.~1, pp. 1--4, Feb. 2023,
  doi: 10.1186/s13040-023-00322-4. [Online]. Available:
  \url{https://www.ncbi.nlm.nih.gov/pmc/articles/PMC9938573/}
\BIBentrySTDinterwordspacing

\end{thebibliography}



\end{document}